\input harvmac

\def\p{\partial}
\def\ap{\alpha'}

\Title{UU-HEP/99-04}{\vbox{\centerline{Holography and 
Noncommutative Yang-Mills}}}

\centerline{Miao Li$^{1,2}$ and Yong-Shi Wu$^{3}$}
\centerline{$^{1}$\it Institute of Theoretical Physics}
\centerline{\it Academia Sinica, P.O. Box 2735}
\centerline{\it Beijing 100080} 
\medskip
\centerline{$^{2}$\it Department of Physics}
\centerline{\it National Taiwan University}
\centerline{\it Taipei 10764, Taiwan}
\centerline{\tt mli@phys.ntu.edu.tw}
\medskip
\centerline{$^{3}$\it Department of Physics}
\centerline{\it University of Utah}
\centerline{\it Salt Lake City, Utah 84112}
\centerline{\it USA}
\centerline{\tt wu@mail.physics.utah.edu}
\bigskip
In this note a lately proposed gravity dual of 
noncommutative Yang-Mills theory is derived 
from the relations, recently suggested by Seiberg 
and Witten, between closed string moduli and 
open string moduli. The only new input one 
needs is a simple form of the running string 
tension as a function of energy. This derivation
provides convincing evidence that string theory 
integrates with the holographical principle, and 
demonstrates a direct link between noncommutative 
Yang-Mills theory and holography.

\Date{September 14, 1999}

\nref\tHooft{G. 't Hooft, ``Dimensional Reduction in Quantum Gravity,''
hep-th/931026.}
\nref\connes{A. Connes. ``{\it Noncommutative Geometry},'' Academic 
Press, Inc.; New York, 1994.}
\nref\suss{L. Susskind, ``The world as a hologram ,'' hep-th/9409089.}
\nref\mald{J.M. Maldacena, ``The Large N Limit of Superconformal 
Field Theories and Supergravity,'' hep-th/9711200.}
\nref\gkpw{E. Witten,  "Anti De Sitter Space and Holography,"
hep-th/9802150; S. Gubser, I. Klebanov and A. Polyakov, 
Phys. Lett. B428 (1998) 105.}
%\nref\gkpw{E. Witten,  "Anti De Sitter Space and Holography,"
%hep-th/9802150; S. Gubser, I. Klebanov and A. Polyakov, 
%Phys. Lett. B428 (1998) 105.}
%\nref\witten{E. Witten, "Anti De Sitter Space and Holography,"
%hep-th/9802150.}
\nref\acr{A. Connes and M. Rieffel,``Yang-Mills For Noncommutative 
Two-Tori,'' in Operator Algebras and Mathematical Physics (Iowa City, 
Iowa, 1985), pp. 237 Contemp. Math. Oper. Alg. Math. Phys. 62, AMS 1987.}
\nref\hw{P.M.~Ho and Y.S.~Wu, ``Noncommutative Geometry and 
$D$-branes,'' Phys. Lett. {\bf B398} (1997) 52, hep-th/9611233.}
\nref\miao{M. Li, "Strings from IIB Matrices", hep-th/9612222.} 
\nref\cds{A. Connes, M. R. Douglas, and A. Schwarz, ``Noncommutative
Geometry and Matrix Theory: Compactification On Tori,'' JHEP {\bf
9802:003} (1998), hep-th/9711162.}
\nref\hwu{P.M.~Ho and Y.S.~Wu, ``Noncommutative Gauge Theories in
Matrix Theory,'' Phys. Rev. {\bf D58} (1998) 066003, hep-th/9801147.}
\nref\dh{M. R. Douglas and C. Hull, ``$D$-Branes And The
Noncommutative Torus,'' JHEP {\bf 9802:008,1998}, hep-th/9711165.}
\nref\mli{M.~Li, ``Comments on Supersymmetric Yang-Mills Theory on a
Noncommutative Torus,'' hep-th/9802052.}
\nref\sw{N. Seiberg and E. Witten, ''String Theory and Noncommutative
Geometry," hep-th/9908142.}
\nref\hi{A. Hashimoto and N. Itzhaki, ``Noncommutative Yang-Mills
and the AdS/CFT correspondence,'' hep-th/9907166.}
\nref\mr{J.M. Maldacena and J.G. Russo, ``Large N Limit of
Non-Commutative Gauge Theories,'' hep-th/9908134.}
\nref\ch{C.-S. Chu and P.-M. Ho,  Nucl. Phys. {\bf B550} (1999) 151, 
hep/th9812219;
V. Schomerus, JHEP {\bf 9906:030} (1999), hep-th/9903205;
F.~Ardalan, H.~Arfaei and M.M.~Sheikh-Jabbari,
JHEP {\bf 02}, 016 (1999) hep-th/981007.}
\nref\ssw{L. Susskind and E. Witten, ``The Holographic Bound in 
Anti-de Sitter Space,'' hep-th/9805114.}
\nref\bs{D. Bigatti and L. Susskind, ``Magnetic Fields, Branes and 
Noncommutative geometry,'' hep-th/9908056.}
\nref\jkt{A. Jevicki, Y. Kazama and T. Yoneya, Phys. Rev. Lett.
81 (1998) 5072. }
\nref\ly{M. Li and T. Yoneya, hep-th/9806240, in a special issue of
"Chaos, Solitons and Fratals," 1998.}

By now it becomes clear that any consistent theory that unifies
quantum mechanics and general relativity requires dramatically 
new ideas beyond what we have been familiar with. Two such 
ideas, {\it the holographic principle \tHooft } and {\it 
noncommutative geometry \connes }, have recently attracted 
increasing attention in string theory community. The holographic 
principle \tHooft, originally motivated by the area dependence of 
black hole entropy, asserts that all information on a quantum theory 
of gravity in a volume is encoded in the boundary surface of the 
volume. Though this principle seems to conflict with our intuition 
from local quantum field theory, string theory as a promising 
candidate of quantum gravity is believed \suss\ to integrate with 
it. Indeed the Maldacena conjecture \mald, motivated by the D-brane 
models of black hole in string theory, is nothing but an embodiment 
\gkpw\ of the holographic principle: There is an equivalence or
correspondence between supergravity (or closed string theory) on an 
anti-de Sitter space, say of five dimensions, and a supersymmetric 
Yang-Mills gauge theory on its four-dimensional boundary. 
 
In a parallel development, Yang-Mills theory on a space with
noncommutative coordinates \acr, which we will call noncommutative 
Yang-Mills theory (NCYM), has been found to arise naturally 
in string theory, first in the multi-D-brane description \hw, 
then in the D-string solution in the IIB matrix model \miao,
then in matrix theory \refs{\cds,\hwu} or string theory \dh\ 
compactifications with nonvanishing antisymmetric tensor 
background, and most recently in a special limit that decouples 
closed string contributions from the open string description for 
coincident D-branes with a constant rank-2 antisymmetric tensor 
B-background (see \sw\ and references therein). 
%(For many papers along these lines, see the references in 
%\sw\. \mcom,snc\ have since followed the original work \cds.) 
Right now it is the last case that is the focus of attention. 
A rather thorough discussion of the aspects of NCYM from the 
open string versus closed string perspectives has been given in 
the work of Seiberg and Witten \sw\ which, among other things, 
also clarifies several puzzles previously encountered in NCYM, 
including the one raised by one of us \mli. Moreover, the 
supposed-to-be gravity duals of NCYM's in the decoupling 
limit, which generalize the usual Maldacena conjecture
without B-background, were also constructed \refs{\hi,\mr}.

One might think that NCYM is relevant only in non-generic 
situations, such as the above-mentioned decoupling limit in 
which the B-background is brought to infinity, so it cannot 
shed much light on deep issues as holography and quantum 
nature of spacetime. In this letter we show that this is 
not true. With an observation made on a direct link between 
the NCYM and its gravity dual, we will try to argue for 
the opposite: Switching on a B-background allows one to 
probe the nature of holography with NCYM, and will probably  
lead to uncovering more, previously unsuspected links 
between a large-N theory and its closed string dual.

One of the central observations in \sw\ is that the natural 
moduli to use in open string theory with ends on a set of N
coincident D-branes in the presence of a constant B-field are 
different from those defined for closed strings. The effective 
action for the system is more elegantly written if one uses 
the open string metric $G_{ij}$ and an antisymmetric tensor 
$\theta^{ij}$, and their relation to the closed string metric 
$g_{ij}$ and the antisymmetric tensor field $B_{ij}$ is
\eqn\rela{\eqalign{G_{ij}&=g_{ij}-(\ap )^2(Bg^{-1}B)_{ij}\cr
\theta^{ij}&=2\pi\ap\left({1\over g+\ap B}\right)_A^{ij},}}
where the subscript $A$ indicates the antisymmetric part. 
Our normalization of the B field differs from that in \sw\ 
by a factor $2\pi$. Seiberg and Witten noted that in the limit 
$\ap\rightarrow 0$ and $g_{ij}\rightarrow 0$ (assuming $B_{ij}$ 
is nondegenerate), it is possible to keep $G_{ij}$ and 
$\theta^{ij}$ fixed with a fixed $B_{ij}$. The tree level 
effective action surviving this limit is the noncommutative 
Yang-Mills action, with a star product of functions defined 
using $\theta^{ij}$:
\eqn\startp{f * g (x)=e^{i/2\theta^{ij}\p^x_i\p^y_j}
f(x)g(y)\mid_{y=x}.} 
This is equivalent to the noncommutativity: 
$x^i*x^j-x^j*x^i=i\theta^{ij}$ \ch.   

The open string coupling constant $G_s$, proportional to the 
Yang-Mills coupling $g^2_{YM}$, is also different from the closed
string coupling constant. The relation between the two is
\eqn\coup{G_s=g_s\left({\det G\over \det(g+\ap B)}\right)^{1/2}.}
It is easy to see that in the ``double scaling limit'' when
$\ap\rightarrow 0$, $g\rightarrow 0$, keeping fixed the open string
coupling $G_s$ (say for D3-branes), $g_s$ must be taken to zero 
too. Thus closed strings decouple from the open string sector 
described by the NCYM. We will see that this perfectly matches 
with the closed string dual description of the NCYM.
 
Using the conventions of \mr, the NS fields in the gravity 
dual proposed for D3-branes with a constant $B_{23}$ are
\refs{\hi,\mr}
\eqn\nsf{\eqalign{ds^2_{str}&=R^2u^2\left[(-dt^2+dx_1^2)
+{1\over 1+a^4u^4}(dx_2^2+dx_3^2)\right] +R^2{du^2\over u^2}
+R^2d\Omega_5^2,\cr
B_{23}&=B{a^4u^4\over 1+a^4u^4},\cr
e^{2\phi}&=g^2{1\over 1+a^4u^4},}}
where we have set $\ap =1$. The constant $B$ is the value of 
$B_{23}$ at the boundary $u=\infty$, and the constant $g$ is 
the closed string coupling in the infrared $u=0$. Here
$R^4=4\pi gN$, and the parameter $a$ is given by
\eqn\para{Ba^2=R^2.}

In addition to the usual $C^{(4)}$ induced by the presence of 
D3-branes, there is also an induced $C^{(2)}$ field. Its presence 
is quite natural for D3-branes.  Recall that a constant $B_{23}$ 
on the branes can be replaced by a constant magnetic field. 
Performing S-duality transformation, this field becomes the 
electric field $E=F_{01}$. This electric field is defined using 
the dual quanta, thus it is equivalent to a $C_{01}$. The 
u-dependent $C_{01}$ is given in \mr: 
\eqn\cfield{C_{01}={a^2R^2\over g}u^4.}

It is natural to interpret the fields appearing in the 
gravity dual \nsf\ as closed string moduli. Note that apart 
from the $u^2$ factor, there is an additional factor 
$1/(1+a^4u^4)$ in the closed string metric on the plane 
$(x_2,x_3)$. Thus if one is to hold the geometry on the plane 
$(t,x_1)$ fixed, then the geometry on the plane $(x_2,x_3)$ 
shrinks when the boundary is approached. By the UV/IR relation \ssw, 
this means that the closed string metric shrinks to zero in the 
UV limit from the open string perspective. Here comes our central 
observation. {\it We identify the UV limit as the ``double scaling 
limit'' of \sw}, thus that $g_{ij}$ shrinks in the UV limit is quite 
natural. In this limit, $\ap$ must also approach zero. This is just 
right in the AdS/CFT correspondence \mald. Note that there is an 
overall factor $R^2u^2$ in \nsf\ for the 4D geometry along D3-branes. 
This redshift factor can be interpreted as the effective string tension
\eqn\teff{\ap_{eff}={1\over R^2u^2}.}
Therefore $\ap_{eff}$ also approaches zero in the UV limit. The 
manner in which it approaches zero compared to $g_{ij}$ agrees with
the limit taken in \sw. Note that here we differ from the philosophy
of \mr\ in which $\ap$ itself is taken to zero, while we have set it
to be $1$. 

Now we are ready to derive the NS fields in the closed string dual
\nsf\ by applying formulas \rela\ and \coup. The way in which
Seiberg and Witten derived these formulas is valid if we treat
strings as effective strings at a fixed energy scale when loop effects
are included. Thus we can take these formulas as giving relations
among the open string moduli and the closed string moduli at
a fixed cut-off $E=u$. Bigatti and Susskind argued \bs\ that in the large
N limit, the effective action of NCYM can be obtained by replacing
the usual product in the effective action of ${\cal N}=4$ SYM
by the star product. This in particular implies that there is no
renormalization for the open string metric, the Yang-Mills coupling
constant $G_s$ and the noncommutative moduli $\theta^{ij}$. 
{\it Now with $\ap$ replaced by $\ap_{eff}$ at a fixed energy scale, 
the closed string moduli are renormalized.} Due to the rotational symmetry
on the $(x_2,x_3)$ plane, we introduce ansatz
\eqn\ans{g_{ij}=f(u)\delta_{ij},\quad B_{ij}=h(u)\epsilon_{ij}.}
The first equation in \rela\ yields
\eqn\feq{\delta_{ij}=\delta_{ij}\left(f+h^2f^{-1}/(R^4u^4)\right),}
or
\eqn\fsl{f^2+{h^2\over R^4u^4}=f.}
The second equation in \rela\ leads to
\eqn\seq{2\pi{1\over R^2u^2}\left(f^2+{h^2\over R^4u^4}\right)^{-1}
{h\over R^2u^2}\epsilon_{ij}=2\pi {a^2\over R^2}\epsilon_{ij},}
where we used the fact that $\theta^{ij}$ is not renormalized and
is given by $(2\pi/B)\epsilon_{ij}$, and $B=R^2/a^2$. This second equation
is just
\eqn\ssl{h=a^2R^2u^4\left(f^2+{h^2\over R^4u^4}\right).}
Combined with eq.\fsl\ we have $h=a^2R^2u^4f$, and substitute this
into eq.\fsl\ we find
\eqn\frl{f(u)={1\over 1+a^4u^4}.}
This is precisely what appeared in \nsf\ which is obtained as a
solution to classical equations of motion in closed string theory.
With $h=a^2R^2u^4f$ we find
\eqn\frls{h(u)={a^2R^2u^4\over 1+a^4u^4}=B{a^4u^4\over 1+a^4u^4},}
also agreeing  with \nsf.

Substitute the above solution into \coup\ with the identification
$G_s=g$, the energy dependent closed string coupling is also
solved
\eqn\cco{g_s=g\left(\det (g+\ap_{eff}B)\right)^{1/2}=g(1+a^4u^4)^{-1/2}.}
Again, this agrees with \nsf. The closed string coupling
becomes weaker and weaker in the UV limit. Although this
UV asymptotic freedom appears in the closed string dual, it does not
mean that there is asymptotic freedom in NCYM, as we have seen
that the Yang-Mills coupling is not renormalized in the large N 
limit.

Having shown that the rather ad hoc looking closed string background
is naturally a solution to \rela\ and \coup, one still has room
to doubt whether this is a coincidence. To check that our procedure is
indeed a correct one, we turn to the case when both $B_{01}$ are
$B_{23}$ are turned on. The solution is also found in \refs{\hi, \mr}.
We use the Euclidean signature for all coordinates. In such a case,
both the geometry of $(t,x_1)$ and the one of $(x_2, x_3)$ shrink
at the boundary in the similar manner. Let $a$ be defined as
before, and $a'$ related to $B_{01}$ in the same way as $a$ is related
to $B_{23}$. We need not to repeat the above steps in deriving
the metric and the B field, since these fields are block digaonalized
and so we will have the similar results. The closed string coupling
is given by, in this case
\eqn\ccou{g_s=g\left(\det(g+\ap_{eff}B)\right)^{1/2}
=g(1+a^4u^4)^{-1/2}(1+a'^4u^4)^{-1/2},}
where the determinant is taken of the matrix including all components.
This result agrees with the classical solution in \refs{\hi,\mr}.
Other part of the closed string metric can not be reproduced so
simply. Due the result of Bigatti and Susskind, and the unbroken
R-symmetry $SO(6)$, it must be identical to that in the AdS case
without B field background.

We see that the relations among the closed string moduli and the
open string moduli contain much more than we could have imagined.
With the input $\ap_{eff}$, they determine the closed string dual
of NCYM! We venture to conjecture that this is a quite general
fact. It can be applied for instance to other Dp-brane case with
constant B field, and perhaps other backgrounds. Also, we expect
that $1/N$ corrections will at least renormalize the Yang-Mills
coupling. The relation \coup\ likely holds in this case, thus
the closed string coupling must be renormalized by the $1/N$
effects. 

%(Insert possible derivation of C here)
 
We now apply our procedure to Dp-branes with constant B field.
Switching on $B_{23}$ only,  
the solution in this case is given in \mr. This solution is similar
to \nsf. For the shrinking metric, we replace $1/(1+a^4u^4)$
by $1/(1+(au)^{7-p})$, where $a$ is determined by $B^2a^{7-p}
=R^{7-p}$. The overall factor $(Ru)^2$ in \nsf\ is replaced by
$(Ru)^{(7-p)/2}$. The u-dependent B field is
\eqn\ubd{B_{23}=B{(au)^{7-p}\over 1+(au)^{7-p}},}
and the dilaton field is
\eqn\dil{e^{2\phi}=g^2u^{(7-p)(p-3)/2}{1\over 1+(au)^{7-p}}.}

To obtain this solution from \rela, we need to use $\ap_{eff}
=(Ru)^{(p-7)/2}$. Introduce the same ansatz for $g_{ij}$ and
$B_{ij}$ as before, the two equations in \rela\ combine to
yield
\eqn\fih{h=(aR)^{(7-p)/2}u^{7-p}f.}
Substitute this into either equation in \rela\ we obtain
\eqn\dpf{f={1\over 1+(au)^{7-p}}.}
This together with \fih\ results in the correct answer for the
B field.

Relation \coup\ determines
\eqn\ccoupl{g^2_s=G^2_sf.}
To agree with \dil\ we must have $G_s^2=g^2u^{(7-p)(p-3)/2}$.
This just means that the open string coupling runs in the same
way as in the case when there is no B field. This fact certainly
agrees with the result of \bs\ in the large N limit.

Our derivation of the closed string dual from NCYM is based 
on the idea that at the dynamical level, namely when the 
quantum effects in NCYM are included by renormalization down 
to fixed energy scale $E=u$, the relations among the open 
string moduli and closed string moduli in view of effective 
strings should be the one derived at the tree level, as done 
in \sw. Although this idea may find its root in some previous 
known facts, such as Fischler-Susskind mechanism, and Polyakov's 
introduction of Liouville field to mimic quantum effects in QCD, 
we have not found a similar precise statement in the literature. 

To complete our derivation of the closed string dual in 
the D3-brane case, we still need to explain the effective 
string tension, namely $\ap_{eff}=1/(R^2u^2)$. As we explained, 
this relation is natural in the supergravity side of the
original AdS/CFT correspondence. We are yet to understand it 
directly in the gauge theory. If we measure energy with 
time $t$ in the gauge theory with metric component $g_{00}=-1$,
we identify $E=u$. In SYM without noncommutative parameters,
this is the only scale, thus $\ap_{eff}\sim 1/E^2=1/u^2$. Yet
we have to attribute the coefficient $1/R^2$ to quantum effects.
Back to NCYM, there are two scales, one of which is determined 
by $\theta$, another energy scale. According to the result of 
\bs, the large N string tension is the same as in ordinary 
SYM, thus $\ap_{eff}=1/(Ru)^2$ in NCYM too.

We have left out the derivation of the metric $du^2/u^2$ in 
the induced dimension. This together with other parts of the 
metric induces an anomaly term in the special conformal 
transformation. This term is computed in SYM in \jkt\ at the 
one loop level, providing a satisfactory interpretation 
of the full AdS metric. Although NCYM is not conformally 
invariant, we expect that a similar calculation to that in 
\jkt\ can be done and thus justifies $du^2/u^2$.

The fact that the geometry on $(x_2,x_3)$ shrinks toward the 
boundary when $B_{23}\ne 0$ has been source of confusions. 
Our scheme clarifies this issue. The geometry in the 
supergravity dual is not to be confused with the geometry 
in NCYM. The latter remains fixed at all energy scales, while 
the geometry felt by closed strings becomes degenerate in 
the UV limit, as it ought to be according to \sw.

There remain quite a few puzzles to be understood within our 
scheme. We need to understand why and how the GKPW prescription 
\gkpw\ works, and why the calculation of the force between a 
pair of heavy quark and anti-quark requires a different 
prescription \mr. Among other things, it can be shown that 
there does not exist a nontrivial geodesic connecting two 
points on the boundary and separated in $x_2$. This already 
indicates interesting behavior of correlation functions at the 
two point function level. We leave these issues for future study.

In conclusion, we have observed that the moduli in the gravity 
duals of NCYM should be understood as closed string moduli, and 
they can be reproduced from the Seiberg-Witten relations \rela\ 
between open and closed string moduli, once the string tension 
in these relations is understood as a running one with simple 
energy dependence as given in \teff. This observation not only 
helps clarify a few puzzles in understanding the gravity dual
of NCYM but also, more importantly, demonstrates a simple and 
direct connection between NCYM and the holographic principle,
either of which is believed to play a role in the ultimate
theoretical structure for quantum gravity. In particular, the 
existence of such a link between NCYM and holography seems to
indicate the fundamental significance of noncommutativity in 
space or spacetime. (For another perspective pointing to the
same direction, see \ly.) To search for appropriate geometric 
framework for the nonperturbative formulation of string/M 
theory, further exploring the connections between these two 
themes should be helpful. For instance, it would be interesting 
to consider further deformation of NCYM, to see what it will 
lead to as the holographic image on boundary for a closed 
string dual in bulk.

%will certainly help us to unravel more and more of the
%ultimate geometric structure. 
%by a further deformation of NCYM that incorporates full closed 
%string effects.

ML acknowledges warm hospitality from Department of Physics,
University of Utah. The work of YSW was supported in part by 
U.S. NSF through grant PHY-9970701. 

\listrefs
\end